\documentclass[sigconf, prologue, dvipsnames]{acmart} 
\AtBeginDocument{%
  }

\setcopyright{rightsretained} 
\copyrightyear{2025}
\acmYear{2025}
\acmDOI{XXXXXXX.XXXXXXX}
\acmConference[EARL '25]{The 2nd EARL Workshop on Evaluating and Applying Recommender Systems with LLMs}{October 5, 2025}{Prague, Czech Republic}
\acmISBN{978-1-4503-XXXX-X/2025/09}

\usepackage{listings}

\begin{document}

\title{HyST: LLM-Powered Hybrid Retrieval over Semi-Structured Tabular Data}

\author{Jiyoon Myung}
\email{jiyoon0424@gmail.com}
\affiliation{%
  \institution{PrompTart LAB, MODULABS}
  \city{Seoul}
  \country{South Korea}
}

\author{Jihyeon Park}
\email{milhaud1201@gmail.com}
\affiliation{%
  \institution{PrompTart LAB, MODULABS}
  \city{Seoul}
  \country{South Korea}
}

\author{Joohyung Han}
\email{ddang8jh@gmail.com}
\affiliation{%
  \institution{PrompTart LAB, MODULABS}
  \city{Seoul}
  \country{South Korea}
}


\begin{abstract}
User queries in real-world recommendation systems often combine structured constraints (e.g., category, attributes) with unstructured preferences (e.g., product descriptions or reviews). We introduce \textbf{HyST} (Hybrid retrieval over Semi-structured Tabular data), a hybrid retrieval framework that combines LLM-powered structured filtering with semantic embedding search to support complex information needs over semi-structured tabular data. HyST extracts attribute-level constraints from natural language using large language models (LLMs) and applies them as metadata filters, while processing the remaining unstructured query components via embedding-based retrieval. Experiments on a semi-structured benchmark show that HyST consistently outperforms tradtional baselines, highlighting the importance of structured filtering in improving retrieval precision, offering a scalable and accurate solution for real-world user queries.
\end{abstract}

\ccsdesc[500]{Information systems~Recommender systems}

\keywords{recommender systems, semi-structured data, large language models, hybrid retrieval, vector database}



\maketitle

\section{Introduction}
Modern recommendation systems increasingly rely on semi-structured data, where items are described by both structured attributes (e.g., category, price, brand) and unstructured content (e.g., descriptions, reviews). Among the various formats of semi-structured data, tabular data is particularly common in practice. Many open recommendation datasets in domains such as business, retail, and travel adopt tabular structures (e.g., CSV) that include attributes like category and brand alongside descriptive fields and user reviews.

User queries in such systems often combine explicit constraints, such as “Italian restaurant in New York” with subjective preferences like “with a cozy atmosphere.” Effectively handling these hybrid queries requires integrating precise filtering over structured fields with semantic understanding of textual descriptions.

Despite the growing capabilities of large language models (LLMs), many retrieval pipelines in recommender systems still struggle with this dual requirement. A common workaround is to flatten all fields into linearized text (e.g., “category: Italian, location: New York, review: cozy atmosphere”) and apply embedding-based retrieval~\cite{oguz-etal-2022-unik}. While simple, this approach often fails to distinguish between hard constraints and soft preferences, which can result in incorrect matches, such as retrieving a French restaurant when the query explicitly asks for Italian.

Other systems attempt to address this challenge by translating user queries into symbolic forms such as SQL, allowing structured conditions to be explicitly executed over relational data. However, these approaches are typically limited to purely structured sources and lack the ability to interpret nuanced preferences conveyed in unstructured fields like reviews or product descriptions. As a result, they often overlook important contextual cues such as user sentiment or subjective attributes, which are essential for personalized recommendation.

To address these limitations, we propose \textbf{HyST}, a unified framework for \textbf{Hy}brid retrieval over \textbf{S}emi-structured \textbf{T}abular data. HyST decomposes a user query into (1) structured filtering conditions and (2) unstructured semantic intent. It uses an LLM to parse the input query and extract structured filters (e.g., category = "Italian", price $<$ \$30), which are then executed using the metadata filtering capabilities of a vector database. The remaining unstructured component (e.g., “cozy atmosphere”) is used for dense retrieval over text fields like descriptions or reviews. This design preserves constraint satisfaction without sacrificing semantic flexibility, while remaining scalable and easy to integrate into existing infrastructure.

We evaluate HyST on a tabular version of the STaRK Amazon benchmark, adapted to simulate semi-structured product recommendation scenarios. Our experiments show that HyST consistently outperforms lexical, dense, and traditional hybrid baselines. Moreover, ablation studies and case analyses confirm the value of LLM-based structured filtering, especially in avoiding retrieval errors caused by purely semantic approaches.

In summary, HyST demonstrates how LLMs can bridge the gap between structured reasoning and semantic relevance in recommendation, offering a practical, scalable solution for interpreting complex user queries over semi-structured data.

\begin{figure*}[t]
  \centering
  \includegraphics[width=\linewidth]{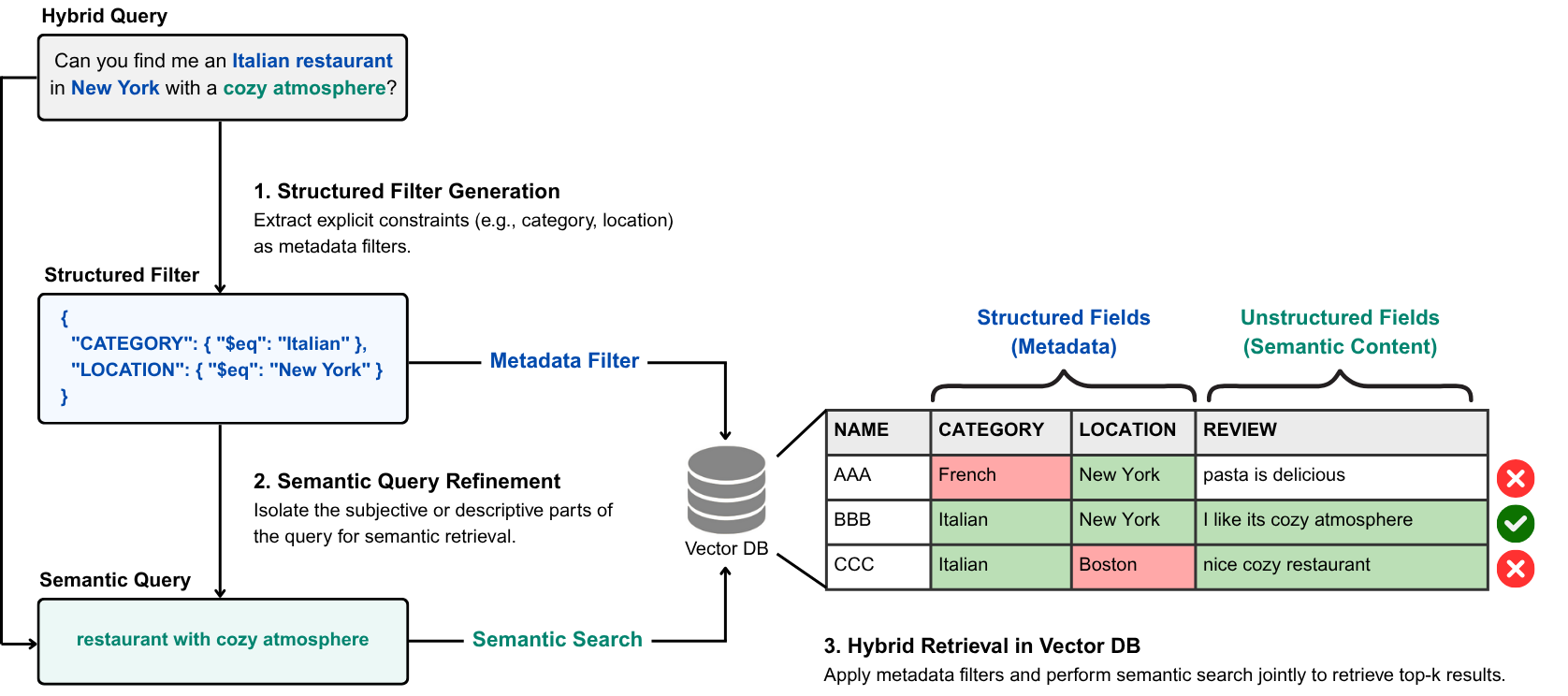}
  \caption{Overview of the HyST framework. Structured constraints are handled via metadata filtering, while unstructured preferences are captured through semantic search.}
  \label{fig:hyst}
\end{figure*}

\section{Related Work}
\label{related_work}

\subsection{Traditional Approaches to Hybrid Retrieval}

A common hybrid strategy in information retrieval combines sparse and dense retrieval models into a unified pipeline. Techniques such as score interpolation~\cite{karpukhin2020densepassageretrievalopendomain} and reciprocal rank fusion (RRF)~\cite{rrf} are often used to aggregate retrieval signals, enhancing robustness across diverse query types by leveraging both lexical precision and semantic generalization.

However, these hybrid models are primarily designed for unstructured text and do not explicitly support structured filtering. The integration of symbolic constraints with neural retrieval remains relatively underexplored.

\subsection{Hybrid Retrieval over Semi-Structured Tabular Data}

Several recent approaches have explored retrieval over semi-structured tabular data that combines structured attributes and unstructured text fields. Some methods flatten both structured and unstructured content into a single textual string and apply embedding-based retrieval~\cite{oguz-etal-2022-unik}. While effective in certain settings, this linearization approach often fails to distinguish between hard constraints and subjective preferences, leading to mismatches in retrieval.

To address this, another line of work extends symbolic query languages with neural operations. For example, recent methods translate user questions into SQL-like queries augmented with LLM-based components that operate over unstructured fields~\cite{liu2024suqlconversationalsearchstructured, biswal2024text2sqlenoughunifyingai}. This allows expressive query compositions combining structured filters with neural answer extraction or summarization. However, these systems often require LLM inference over many candidate rows at runtime, which limits scalability.

\subsection{LLM-Driven Metadata Filtering for Retrieval}

Beyond SQL-like symbolic queries, recent work has explored using LLMs to generate structured filters that can be directly applied in retrieval pipelines. For instance, ~\citet{Poliakov_2025} show that LLMs can generate filter expressions compatible with vector databases, allowing natural language queries to be constrained by metadata. However, much of this prior work emphasizes coarse-grained filtering (e.g., filtering by domain, source) and does not extend to the fine-grained attribute-level filtering often required in recommendation systems.

\begin{table*}[h]
\centering
\caption{Examples of selected queries requiring both structured filtering (\textcolor{blue}{blue}) and semantic reasoning (\textcolor{teal}{teal}).}
\label{tab:query_examples}
\begin{tabular}{@{}c|l@{}}
\toprule
\textbf{Index} & \textbf{Query} \\
\midrule
1540 & Can you suggest a \textcolor{teal}{high-quality} \textcolor{blue}{fishing line} from \textcolor{blue}{Sufix} that offers \textcolor{teal}{good value for money}? \\
2573 & Is there a \textcolor{teal}{user-friendly and highly precise} \textcolor{blue}{hunting bow} from \textcolor{blue}{Martin} that you could suggest? \\
7118 & Can you recommend an \textcolor{blue}{ORCA brand} \textcolor{blue}{swim cap} that's \textcolor{teal}{ideal for training or racing purposes}? \\
7714 & What are some \textcolor{teal}{cute design options} for women's capri \textcolor{blue}{leggings} from \textcolor{blue}{Zumba} \textcolor{teal}{made of polyester and spandex}? \\
\bottomrule
\end{tabular}
\end{table*}

\section{Methodology: HyST Framework}

We propose \textbf{HyST}, a hybrid retrieval framework that leverages large language models (LLMs) to interpret natural language queries and perform precise, scalable retrieval over semi-structured data. HyST decomposes the user query into structured and unstructured components and executes both types of constraints within a vector database using native metadata filtering and semantic search.

As illustrated in Figure~\ref{fig:hyst}, the pipeline consists of three stages: (1) metadata filter generation, (2) query refinement, and (3) hybrid retrieval using a vector database.

\subsection{Metadata Filter Generation}

In the first stage, HyST uses an LLM to extract structured constraints from the user's query. These typically correspond to well-defined attributes such as category, price, location, brand, or availability. The LLM is prompted to generate structured filter conditions in the format accepted by vector databases (e.g., Pinecone), which typically follow a JSON-like schema using operators such as `\$eq` (equal), `\$in` (list membership), and `\$lt` (less than).

For example, given the query \textit{"Show me Italian or French restaurants in New York with a cozy atmosphere, priced under \$100, and with parking available"}, the model produces:

\begin{lstlisting}[frame=single, backgroundcolor=\color{gray!20}, basicstyle=\ttfamily\scriptsize]
{
  "CATEGORY": {"$in": ["Italian", "French"]},
  "LOCATION": {"$eq": "New York"},
  "PRICE": {"$lt": 100},
  "PARKING": {"$eq": "Y"}
}
\end{lstlisting}

This metadata filter is passed directly to the vector database, which uses it to constrain the candidate set before similarity search is performed.

\subsection{Query Refinement for Semantic Search}

Next, HyST isolates the unstructured, subjective aspects of the query such as ambiance, tone, or style which cannot be captured by structured fields. The LLM is prompted to remove the portions already encoded in the structured filter and output a concise semantic query for downstream embedding-based retrieval.

In the above example, the phrase \textit{"restaurant with cozy atmosphere"} remains as the refined semantic query. This enables the system to focus vector search on aspects that are context-sensitive and not explicitly structured in the schema.

\subsection{Hybrid Retrieval over Vector Database}

In the final stage, HyST performs hybrid retrieval using a vector database that supports both metadata filtering and semantic similarity search. Each item in the database contains both structured metadata (used for filtering) and a dense embedding of unstructured content (used for vector search).

At query time, the database first filters records using the LLM-generated metadata conditions, then performs semantic similarity search over the embeddings of the remaining candidates using the refined query. This enables precise enforcement of hard constraints while also ranking results by their relevance to the user's implicit preferences.

By integrating both types of signals within a unified infrastructure, HyST achieves both constraint satisfaction and semantic flexibility, without requiring external post-filtering or custom execution logic.

\section{Experiments}

\subsection{Dataset}

We employ the STaRK Amazon benchmark~\cite{wu2024starkbenchmarkingllmretrieval}, which provides a combination of semi-structured knowledge bases and a QA dataset, making it well-suited for evaluating hybrid retrieval methods. Originally designed for product recommendation systems, STaRK Amazon features a product knowledge base that highlights customer-centric attributes such as product quality, functionality, and style. Complementing this, the QA dataset contains natural, conversational queries that realistically reflect user interactions with product search systems.

While there exist open-domain recommendation datasets in tabular format (e.g., Yelp Open Dataset), these typically lack associated QA annotations, making it difficult to perform quantitative evaluation of retrieval performance. In contrast, STaRK provides a rich set of question-answer pairs aligned with a semi-structured product catalog, enabling precise benchmarking. To focus our evaluation on tabular-style semi-structured data rather than graph-based knowledge representations, we transform the original STaRK multi-relational schema into a flat table format.

Specifically, we extract five fields from the original knowledge base: \textit{title}, \textit{brand}, \textit{category}, \textit{description}, and \textit{reviews}. Among these, \textit{brand} and \textit{category} are used as structured attributes and stored as metadata in the vector database, enabling fast and precise filtering. The remaining fields—\textit{title}, \textit{description}, and \textit{reviews}—are concatenated into a single unstructured textual input, embedded using a dense encoder, and stored as content for semantic retrieval. This flattened schema allows HyST to combine LLM-driven metadata filtering and semantic vector search in a unified and efficient retrieval pipeline.

One notable limitation of the STaRK dataset is the restricted availability of structured fields. While the dataset contains additional attributes, such as color, we found that questions specifically querying these aspects were sparse. As a result, we primarily relied on category and brand as structured fields, which constrained the diversity of structured queries we could evaluate. This limitation made it challenging to assess scenarios involving numerical comparisons, like price ranges or ratings. Despite these constraints, STaRK remained the most suitable option due to its QA dataset, as other semi-structured datasets lacked similar question-answer pairs, making them less applicable to our research focus.

To create a realistic evaluation setup, we manually select 76 representative queries from the QA dataset. These queries are chosen because they require both structured filtering (e.g., by brand, category) and semantic reasoning (e.g., user sentiment, product features). Queries requiring multi-hop relations like \textit{also\_bought} or \textit{also\_viewed} are excluded, as they fall outside the scope of our tabular framework. Table~\ref{tab:query_examples} shows examples of the selected hybrid queries.

To build a focused yet challenging evaluation set, we included only products that either appeared as ground-truth answers to the selected queries or matched the brands or categories specified within them. This filtering process significantly reduced the dataset from over 1 million entities to a more manageable set of 3,335 products. Despite the reduction, this curated subset preserves the retrieval difficulty while enabling more efficient evaluation. By employing this approach, we create a benchmark that aligns well with the semi-structured retrieval framework of HyST, allowing for a robust assessment of its ability to combine structured filtering with semantic ranking.

\subsection{Implementation Details}

We implement HyST using the following components:
\begin{itemize}
    \item \textbf{Vector Database:}  We utilize Pinecone as the vector database to efficiently perform similarity-based ranking on the metadata-filtered results, employing cosine similarity as the distance metric.

    \item \textbf{Embedding Model:} We employ OpenAI’s text-embedding-3-small model to create dense vector embeddings (dimension = 1536) for both the refined user queries and the unstructured text fields (description, reviews) associated with each product.

    \item \textbf{LLM:} To dynamically generate metadata filtering conditions from natural language queries, we leverage GPT-4o (via OpenAI API). The LLM is specifically prompted to identify structured components (such as brand, category) and transform them into a Pinecone-compatible filtering format. During the generation of filtering conditions, we use a temperature setting of 0.3 and top-p of 0.8. The complete prompt used for generating these conditions is detailed in Appendix \ref{lst:prompt}.
\end{itemize}

\subsection{Baselines}
To assess the effectiveness of HyST, we compare it against a set of representative baselines that reflect different paradigms in information retrieval. Unlike HyST, these baselines do not utilize metadata as a distinct channel for structured filtering. Instead, all available fields are linearized into a flat text string in the format \textit{brand: value, category: value, description: value}, and embedded as a whole.

\begin{itemize}

\item \textbf{BM25 (Lexical)}~\cite{robertson1995bm25}: A classical sparse retrieval method based on term frequency-inverse document frequency (TF-IDF). It ranks documents using exact keyword matches and serves as a strong lexical baseline, particularly for structured or keyword-centric queries.

\item \textbf{DPR (Dense)}~\cite{karpukhin2020dpr}: A dense retrieval model that encodes queries and documents separately using dual encoders. We use the \textit{facebook/dpr-ctx\_encoder-single-nq-base} model, retrieving documents based on cosine similarity in the embedding space.

\item \textbf{ANCE (Dense)}~\cite{xiong2020approximatenearestneighbornegative}: A dense retriever trained with dynamic hard negatives through Approximate Nearest Contrastive Estimation. We use the pretrained \textit{castorini/ance-msmarco-passage} model for evaluation.

\item \textbf{ColBERTv2 (Dense)}~\cite{santhanam2022colbertv2effectiveefficientretrieval}: A late interaction retrieval model that computes fine-grained token-level similarity between queries and documents. We use the pretrained \textit{colbert-ir/} \textit{colbertv2.0} model.

\item \textbf{BM25 + DPR (Hybrid)}~\cite{karpukhin2020densepassageretrievalopendomain}: A hybrid pipeline that combines BM25 and DPR scores via weighted sum. This two-stage approach integrates sparse and dense signals, with $\lambda=0.5$ controlling their contribution.

\item \textbf{BM25 + OpenAI Embeddings (Hybrid)}: Similar to the previous hybrid setting, but replaces DPR with OpenAI’s \textit{text-embedding-3-small} model for dense retrieval. This evaluates performance using a state-of-the-art embedding model.

\item \textbf{Linearized Semantic Retrieval}~\cite{oguz-etal-2022-unik}: 
A purely semantic baseline where both structured attributes and unstructured fields are flattened into a single textual string (e.g., "category: socks, brand: Nike, review: very durable") before encoding. To ensure comparability, we use the same embedding model (OpenAI’s \textit{text-embedding-3-small}) and vector database (Pinecone) as in HyST. Retrieval is performed using vector similarity without any structured filtering or query decomposition.

\end{itemize}

\subsection{Evaluation Metrics}

The performance of HyST is measured using standard retrieval metrics commonly employed in recommendation and retrieval tasks. These metrics align with the evaluation strategy used in the STaRK benchmark while offering a more nuanced assessment by focusing on precision rather than simple hit rates.
\begin{itemize}
    \item \textbf{Precision@k:} Measures the proportion of relevant items among the top-$k$ results returned by the model. We use $k = 1, 5, 10$ for evaluation.
    
    \item \textbf{Recall@k:} Quantifies the proportion of relevant items identified within the top-$k$ results, indicating how well the model captures all possible relevant answers. We set $k = 20$ following the STaRK benchmark design, as the number of relevant items per query does not exceed 20.
    
    \item \textbf{Mean Reciprocal Rank (MRR):} Computes the average of the reciprocal ranks of the first relevant item across all queries. This metric emphasizes the position of the first correct answer, which is especially important in recommendation settings where users are more likely to engage with top-ranked results.
\end{itemize}

\section{Result}
\label{result}

\begin{table*}[!h]
\centering
\captionsetup{justification=centering}
\caption{Retrieval performance comparison of HyST and baseline methods. The best score for each metric is highlighted in \textbf{bold}.}
\label{tab:retrieval_metrics}
\begin{tabular}{l|ccccc}
\toprule
\textbf{Method} & \textbf{Precision@1} & \textbf{Precision@5} & \textbf{Precision@10} & \textbf{Recall@20}  & \textbf{MRR} \\
\midrule
BM25~\cite{robertson1995bm25} & 0.8026 & 0.6053 & 0.5105 & 0.6960 & 0.8564   \\
DPR~\cite{karpukhin2020dpr} & 0.3816 & 0.3605 & 0.3105 & 0.4037 & 0.5283\\
ANCE~\cite{xiong2020approximatenearestneighbornegative} & 0.8026 & 0.6237 & 0.5303 & 0.7352 & 0.8639  \\
ColBERTv2~\cite{santhanam2022colbertv2effectiveefficientretrieval} & 0.6316 & 0.4842 & 0.4013 & 0.5651 & 0.7224  \\
BM25 + DPR~\cite{karpukhin2020densepassageretrievalopendomain} & 0.8026 & 0.6263 & 0.5566 & 0.7740 & 0.8604  \\
BM25 + OpenAI Embeddings & 0.8289 & 0.6526 & 0.5803 & 0.8040 & 0.8881  \\
Linearized Semantic Retrieval~\cite{oguz-etal-2022-unik} & 0.8947 & 0.6947 & 0.5974 & 0.8019 & 0.9232 \\
\textbf{HyST (Ours)} & \textbf{0.9211} & \textbf{0.8349} & \textbf{0.8022} & \textbf{0.8063} & \textbf{0.9265} \\
\bottomrule
\end{tabular}
\end{table*}

\subsection{Overall Performance}

Table~\ref{tab:retrieval_metrics} reports the performance of HyST compared to various lexical, dense, and hybrid baselines. HyST achieves the best results on all metrics: precision @ 1, precision @ 5, precision @ 10, recall @ 20, and mean reciprocal rank (MRR) — demonstrating its strong effectiveness in handling semi-structured recommendation queries. Notably, HyST outperforms the widely used linearized semantic baseline~\cite{oguz-etal-2022-unik} by a significant margin, especially in top-ranked precision metrics (e.g., +2.6 points in P@1 and +14.0 points in P@5), highlighting the benefit of explicitly separating structured and unstructured query components.

Among the baselines, hybrid methods combining sparse and dense retrieval (e.g., BM25 + OpenAI Embeddings) show improved robustness over pure dense models like DPR and ColBERTv2. Interestingly, the linearized approach also performs well, likely due to the use of strong embedding models and natural alignment between text and query. However, its lack of structural filtering results in lower precision at deeper ranks (e.g., P@10 and R@20), where constraint violations become more prominent. HyST overcomes this limitation by incorporating LLM-generated filters that precisely narrow down candidates before semantic ranking.

These results underscore the importance of structured filtering in LLM-powered recommendation pipelines, and validate HyST's hybrid design as a scalable and accurate solution for complex real-world queries.

\begin{table}
\centering
\small  
\caption{Ablation study on the effect of query refinement in HyST.}
\label{tab:ablation_query_refinement}
\begin{tabular}{l|cccc}
\toprule
\textbf{Variant} & \textbf{P@1} & \textbf{P@5} & \textbf{P@10} & \textbf{R@20} \\
\midrule
HyST w/o query refinement & \textbf{0.9211} & \textbf{0.8586} & \textbf{0.8167} & 0.8046 \\
HyST (full) & \textbf{0.9211} & 0.8349 & 0.8022 & \textbf{0.8063} \\
\bottomrule
\end{tabular}
\end{table}

\subsection{Ablation Study: Impact of Query Refinement}

We conduct an ablation study to assess the role of query refinement in HyST. As shown in Table~\ref{tab:ablation_query_refinement}, removing the refinement step—i.e., using the full original query for dense retrieval rather than isolating the unstructured component—results in slightly higher performance in \textbf{Precision@5} (0.8586 vs. 0.8349) and \textbf{Precision@10} (0.8167 vs. 0.8022), while overall Precision@1 and Recall@20 remain comparable.

This finding suggests that query refinement, though conceptually appealing for disentangling hard constraints from soft preferences, does not always lead to better retrieval in practice. In datasets like STaRK, unstructured fields (e.g., product titles, descriptions, reviews) often contain rich signals that overlap with structured attributes—such as brand names, categories, or specifications. As a result, refining the query to isolate only the subjective component (e.g., turning “durable Adidas socks” into “durable”) may inadvertently discard important cues embedded in the original phrasing.

These results indicate that the benefit of query refinement is highly dependent on how structured information is distributed across textual fields. For datasets where unstructured content already encodes rich, structured cues, retaining the full query may better preserve user intent. Accordingly, we recommend treating query refinement as an optional step that can be toggled based on the characteristics of the dataset and the presence of redundant structured signals in unstructured fields.

\subsection{Case Study: HyST vs. Semantic Search Only}

To better understand HyST’s advantage over purely semantic methods, we analyzed cases where HyST correctly retrieved the target items while Linearized Semantic Retrieval failed. These cases, summarized in Table~\ref{tab:case_study_multiple}, are selected to analyze the typical failure modes of semantic-only retrieval and highlight the core advantages of combining structured filtering with semantic ranking.

A common source of error in Semantic Search Only is its inability to enforce exact structured constraints, such as brand or category. For example, when the query asks for a Spyder brand paintball marker, Semantic Search Only retrieves a product from a different brand (3Skull) that is semantically similar, while HyST correctly restricts candidates to those with the specified brand. 

These examples clearly demonstrate that semantic similarity is insufficient when user queries contain explicit structured filters. By leveraging LLM-generated metadata conditions and applying them within the vector database, HyST ensures that only candidates satisfying the structured constraints are considered for semantic ranking. This hybrid strategy effectively reduces irrelevant results and improves retrieval accuracy in real-world recommendation scenarios.

\subsection{Error Analysis}

Although HyST achieves high performance across standard metrics, analyzing its failure cases reveals practical challenges that affect specific query types and retrieval stages.

On the structured side, the LLM is generally effective at extracting accurate metadata filters from natural language queries. In our evaluation on the STaRK dataset, where queries are relatively simple and schema-aligned, such mistakes were extremely rare. Only one instance was observed in which the LLM hallucinated a non-existent attribute (e.g., \texttt{"DATA\_TIMELINE"}). Although this kind of error can often be mitigated through prompt refinement or schema-aware constraints, it highlights a potential risk when handling more complex or open-ended queries.

In contrast, most retrieval errors stemmed from the semantic search component. In these cases, the retrieved items satisfied the structured filters but failed to capture the nuanced, subjective intent of the query, such as tone, sentiment, or descriptive qualities. As HyST relies on a fixed embedding model for this stage, retrieval accuracy is inherently limited by the model’s semantic expressiveness.

\begin{table*}[!h]
\centering
\captionsetup{justification=centering}
\caption{
Case Study: Comparison between HyST and Semantic Search Only. 
\textcolor{blue}{Blue} highlights structured constraints (e.g., brand, category) mentioned in the query. 
\textcolor{ForestGreen}{Green} highlights correctly matched structured values in the retrieved results. 
\textcolor{red}{Red} highlights constraint violations in incorrect results.
}
\label{tab:case_study_multiple}
\begin{tabular}{p{4.5cm}|p{5.5cm}|p{5.5cm}}
\toprule
\textbf{Query} & \textbf{HyST} & \textbf{Linearized Semantic Retrieval} \\
\midrule
What is a highly rated lightweight \textcolor{blue}{paintball} marker with great firepower from the \textcolor{blue}{Spyder} brand, as recommended by reviewers?
&
\textbf{TITLE:} Kingman Spyder 08 Pilot Paintball Marker/Gun - Black \newline
\textbf{BRAND:} \textcolor{ForestGreen}{Spyder} \newline
\textbf{CATEGORY:} \textcolor{ForestGreen}{paintball}, complete paintball sets\newline
\textbf{REVIEW:} ...This gun is the best I've ever played with...
&
\textbf{TITLE:} Spyder MR100 Pro Paintball Marker Gun 3Skull Sniper Set \newline
\textbf{BRAND:} \textcolor{red}{3Skull} \newline
\textbf{CATEGORY:} \textcolor{ForestGreen}{paintball}, complete paintball sets \newline
\textbf{REVIEW:} ...Great marker, and a good price for all the gear you need to start playing today....
\\ \midrule
Is there a \textcolor{blue}{table tennis} table by \textcolor{blue}{Halex} that you would recommend?
&
\textbf{TITLE:} Halex Sigma Table Tennis Table\newline
\textbf{BRAND:} \textcolor{ForestGreen}{Halex} \newline
\textbf{CATEGORY:} \textcolor{ForestGreen}{table tennis} \newline
\textbf{REVIEW:} ...We are happy  overall with the table...
&
\textbf{TITLE:} Halex Washer Toss Target Outdoor Games \newline
\textbf{BRAND:} \textcolor{ForestGreen}{Halex} \newline
\textbf{CATEGORY:} \textcolor{red}{toss games} \newline
\textbf{REVIEW:} ...The game itself looks nice....
\\ \midrule
Is there a user-friendly and highly precise \textcolor{blue}{hunting bow} from \textcolor{blue}{Martin} that you could suggest?
&
\textbf{TITLE:} Martin Archery J8S Fletching Jig Straight\newline
\textbf{BRAND:} \textcolor{Green}{Martin} \newline
\textbf{CATEGORY:} \textcolor{Green}{archery}\newline
\textbf{REVIEW:} ...This jig is great...
&
\textbf{TITLE:} Trophy Ridge Speed Comp XV525IR Crossbow Scope\newline
\textbf{BRAND:} \textcolor{red}{Bear Archery} \newline
\textbf{CATEGORY:} \textcolor{Green}{archery}\newline
\textbf{REVIEW:} ...I think this is a good product, ...
\\
\bottomrule
\end{tabular}
\end{table*}

\section{Limitations}

While HyST achieves promising results on current benchmarks, its design and evaluation still leave open areas for further exploration and improvement.

First, although HyST’s modular architecture enables transparent reasoning by decoupling structured filtering and semantic search, its performance remains closely tied to the capabilities of its underlying components. In particular, the effectiveness of structured filtering depends on the LLM's ability to interpret queries and generate accurate filters, while semantic retrieval quality hinges on the expressiveness of the embedding model. Enhancing either component, such as through schema-aware prompting or domain-specific embeddings, could further improve robustness to diverse and ambiguous queries.

Second, the evaluation is constrained to relatively simple filtering logic. The STaRK dataset does not include queries requiring numeric range filtering, nested conditions (e.g., “under \$100 but above \$50”), or aggregation-based reasoning (e.g., “products with more than 100 reviews”). As a result, HyST’s ability to handle complex filtering remains untested. These types of queries are nevertheless common in practical recommendation scenarios and deserve further attention.

Third, the schema fields used in evaluation, particularly category and brand, feature hundreds of fine-grained values with overlapping semantics. LLMs may struggle to map paraphrased or informal query terms to exact schema entries. To reduce noise, we limited the evaluation to a smaller, cleaner subset of these values. While this improves experimental consistency, it also reduces realism. Future systems must address schema alignment challenges at scale through more robust entity linking or controlled vocabularies.

Future research could address these challenges by developing or extending benchmarks that include more diverse semi-structured queries, especially those involving numeric filtering or multi-condition logic. Incorporating a broader range of query types would enable a more comprehensive evaluation of hybrid retrieval frameworks like HyST, particularly in applications requiring advanced filtering and complex reasoning.

\section{Conclusion}

We presented HyST, a hybrid retrieval framework for semi-structured data that combines large language model (LLM)-driven query understanding with the native filtering and dense retrieval capabilities of vector databases. HyST bridges structured metadata constraints and unstructured semantic relevance by dynamically extracting fine-grained filters from natural language queries and applying them within the vector search pipeline.

We evaluated HyST on the STaRK Amazon benchmark, which we adapted into a tabular format to align with the semi-structured data retrieval task. Our experiments demonstrated that HyST consistently outperforms traditional lexical, dense, and hybrid baselines, highlighting the effectiveness of combining LLM-based structured filtering with vector-based semantic search.

By leveraging the native filtering and dense retrieval capabilities of vector databases, HyST offers a practical and scalable solution for real-world recommendation systems, where both categorical precision and contextual relevance are essential.

\begin{acks}
This research was supported by Brian Impact Foundation, a non-profit organization dedicated to the advancement of science and technology for all.
\end{acks}

\bibliographystyle{ACM-Reference-Format}
\bibliography{hyst}

\clearpage
\onecolumn
\appendix

\section{LLM Prompt for Metadata Filtering Generation}
\label{app:prompt}

\begin{lstlisting}[frame=single, breaklines=true, backgroundcolor=\color{gray!20}, basicstyle=\ttfamily\footnotesize, caption={LLM Prompt for Generating Metadata Filters}, label={lst:prompt}]
You are an expert data retrieval assistant. Your task is to generate structured metadata filtering conditions from natural language queries. 
You will identify key attributes, categories, and constraints in the user input and convert them into a metadata filter format compatible with vector databases like Pinecone.

### Schema Information:
- Table name: AMAZON
- Columns:
    - BRAND (single): Brand name of the product; allowable values: {allowable_brands}
    - CATEGORY (multiple): Product classification categories, multiple categories separated by commas; allowable values: {allowable_categories}

### Guidelines:
1. Identify structured components such as category, location, brand, price, or other attributes based on the given schema.
2. Use appropriate comparison operators:
   - "$eq" for exact matches when the column type is "single" (e.g., brand).
   - "$in" for multiple options when the column type is "multiple" or the query mentions a list of possible values.
   - "$lt" for less than comparisons for numerical values (e.g., price under 100).
   - "$gt" for greater than (e.g., rating above 4).
   - "$between" for range conditions (e.g., price between 50 and 100).
3. If the query contains multiple conditions, combine them using AND logic.
4. Always use the "$in" operator for columns specified as "multiple" in the schema, even if the input has only one value.
5. Ignore subjective or descriptive language that cannot be directly used for structured filtering.
6. Always match the categorical values strictly with the allowable values specified in the schema.
7. Answer only the filtering conditions in JSON format.

### Example Queries:
Query 1:
Find cozy Italian or French restaurants in New York with prices under $100.
Expected Output:
{
    "CATEGORY": {"$in": ["Italian", "French"]},
    "LOCATION": {"$eq": "New York},
    "PRICE": {"$lt": 100}
}

Query 2:
Show me electronics from Sony or Samsung that cost between $300 and $500.
Expected Output:
{
    "CATEGORY": {"$eq": "Electronics"},
    "BRAND": {"$in": ["Sony", "Samsung"]},
    "PRICE": {"$between": [300, 500]}
}

Query 3:
Recommend luxury hotels in Paris with a rating above 4.5.
Expected Output:
{
    "CATEGORY": {"$eq": "Luxury Hotel"},
    "LOCATION": {"$eq": "Paris"},
    "RATING": {"$gt": 4.5}
}

User question: {question}
\end{lstlisting}

\end{document}